\documentclass[a4paper]{JHEP3}

\usepackage{epsfig,multicol,bbm}
\usepackage{amsmath}
\usepackage{amssymb}
\usepackage{amsfonts}
\usepackage{graphicx,epsfig}
\usepackage{subfigure}
\usepackage{exscale}
\usepackage{float}
\usepackage[numbers,sort&compress]{natbib}

\newcommand\fverb{\setbox\fverbbox=\hbox\bgroup\verb}
\newcommand\fverbdo{\egroup\medskip\noindent%
			\fbox{\unhbox\fverbbox}\ }
\newcommand\fverbit{\egroup\item[\fbox{\unhbox\fverbbox}]}
\newbox\fverbbox

\newcommand{\h}{\vec h}
\newcommand{\e}{\mbox{e}}
\newcommand\eq{\begin{equation}}
\newcommand\en{\end{equation}}
\newcommand{\ket}{\rangle}
\newcommand{\bra}{\langle}

\title{The width of the color flux tube at 2-Loop order}

\author{F.\ Gliozzi\\
Dipartimento di Fisica Teorica, Universit\`a di Torino, and \\
INFN, Sezione di Torino, via P.\ Giuria 1, 10125 Torino, Italy. \\
	E-mail: \email{gliozzi@to.infn.it}}

\author{M.\ Pepe\\
INFN, Sezione di Milano-Bicocca\\ 
Edificio U2, Piazza della Scienza 3, 20126 Milano, Italy. \\
	E-mail: \email{pepe@mib.infn.it}}

\author{U.-J.\ Wiese\\
Albert Einstein Center for Fundamental Physics \\
Institute for Theoretical Physics, Bern University\\
Sidlerstr.\ 5, 3012 Bern, Switzerland.\\
	E-mail: \email{wiese@itp.unibe.ch}}

\abstract{The color flux tube connecting a static quark-anti-quark pair in Yang-Mills
  theory supports massless transverse fluctuations, which are the Goldstone bosons of
  spontaneously broken translation invariance. Just as in chiral perturbation theory, the
  dynamics of these Goldstone bosons is described by a systematic low-energy effective
  field theory. We use the effective theory to calculate the width of the fluctuating
  string at the 2-loop level, using both cylindrical and toroidal boundary conditions. At
  zero temperature, the string width diverges logarithmically with the quark-anti-quark
  distance $r$. On the other hand, at low but non-zero temperature $T = 1/\beta$, for $r
  \gg \beta$ the string width diverges linearly.}


\begin{document} 

\section{Introduction}

Non-perturbative phenomena emerging in strongly coupled systems are quite 
common in physics. For example, in condensed matter physics, strongly coupled 
electrons are responsible for quantum antiferromagnetism and high-temperature 
superconductivity, whose origin remains mysterious even after decades of 
intensive research. In particle physics, the confinement of quarks is one of 
the most fundamental features of the strong interactions, and our understanding 
of the non-perturbative dynamics underlying quark confinement is still far from
being complete. In general, it is very challenging to understand the dynamical 
mechanisms that are responsible for non-perturbative effects, also because
collective modes play a crucial role. 

However, even if the origin of some non-perturbative effect remains unclear, 
one may be able to quantitatively understand some of its consequences. In 
particular, one may analyze the symmetries --- internal or related to space-time
--- of the ground state of a non-perturbative system, and then work out the 
consequences of spontaneous symmetry breaking, even if the dynamics leading to 
this phenomenon may remain unclear. In this framework, the typical energy scale 
of the non-perturbative phenomenon represents an upper bound for the energy of 
the phenomena that one wants to investigate. When a continuous symmetry breaks
spontaneously, the resulting Goldstone bosons are the natural degrees of 
freedom to be taken into account in the low-energy regime of the theory. The
Lagrangian of the corresponding systematic low-energy effective theory consists
of all terms respecting the internal and space-time symmetries of the system. 
The terms in the Lagrangian are multiplied by low-energy parameters, whose
explicit values can be derived only from the underlying quantum theory. They 
represent the high-energy non-perturbative input for the low-energy dynamics.

The effective field theory approach is a very powerful tool for investigating 
strongly coupled systems. In particular, it provides a way to perform analytic 
calculations in physical systems that would otherwise be unmanageable. For 
example, chiral Lagrangians describe pion and pion-nucleon systems in a
very accurate manner. Similarly, a systematic low-energy theory for magnons 
provides an excellent description of the low-energy dynamics of quantum 
antiferromagnets. 

In 1980,  L\"uscher, Symanzik, and Weisz \cite{Lue80} proposed a low-energy 
effective string description for the color flux tube connecting a static 
quark-anti-quark pair in the confined phase of $d$-dimensional Yang-Mills 
theory. During its time-evolution, the flux string sweeps out a 2-dimensional 
world-sheet, thereby spontaneously breaking the translation invariance in the 
transverse directions. The $(d-2)$ Goldstone bosons resulting from that 
breaking are the relevant degrees of freedom in the low-energy effective 
description of the string dynamics. The leading term of the effective action 
describes the dynamics of a thin string in the free-field approximation. Two 
non-trivial results follow from this observation. First, the leading correction 
to the linear term in the static potential \cite{Lue81} is $-\pi (d-2)/(24 r)$, 
where $r$ is the distance between the two static sources. Second, the flux tube 
width is not fixed but increases like $(d-2)\log(r/r_0)/(2 \pi \sigma)$
\cite{Lue81a}, with the string tension $\sigma$ and a length scale $r_0$ 
entering the effective theory as low-energy parameters.

Many numerical lattice simulations have successfully demonstrated the validity 
of the predictions of the effective theory 
\cite{has,cas96,Cas96a,Luc01,Jug02,Lue02,Cas04,Cas06,Har06,giu07,Bri08,Ath08,All08,Bra09,gpw,Bak10}. However, the very high numerical accuracy 
of the Monte Carlo simulations calls for improving the analytic results of the 
low-energy effective description. The sub-leading corrections to the free-field 
action of the $(d-2)$ Goldstone bosons depend on a number of parameters that, 
in general, can be fixed only by matching to the underlying Yang-Mills theory. 
Remarkably, L\"uscher and Weisz have pointed out that the low-energy effective 
string theory has a symmetry that had been overlooked before~\cite{Lue04}. This 
new symmetry implies some relations between the values of the low-energy 
parameters of the effective theory. Interestingly, the Nambu-Goto action 
satisfies the constraints that one finds at next-to-leading order. In the 
special case of $d=3$, the values of the low-energy parameters are completely 
fixed up to the next-to-next-to-leading order and they turn out to be the same 
as those of the Nambu-Goto action \cite{aha09}. In this paper we report on the 
computation of the correction to the  increase of the flux tube 
width resulting from the sub-leading correction to the free-field action in $d$ 
dimensions. The final expressions involve the Dedekind $\eta$ function and the
Eisenstein series $E_2$ and $E_4$. Due to the modular inversion property
of those functions, we have the next-to-leading correction to the string
width both at zero and at finite (but low) temperature.

The paper is organized as follows. Section 2 contains the description of the systematic
low-energy effective string theory for the dynamics of the confining string. In section 3
toroidal boundary conditions, which describe a closed string wrapping around a compact
spatial dimension, are investigated.  Similarly, section 4 discusses cylindrical boundary
conditions, which correspond to the propagation of an open string that ends in the static
quark-anti-quark charges. For both boundary conditions the width of the string is
analytically calculated at the 2-loop level. We present our conclusions in section
5. Technical details are described in four appendices.

\section{Effective string theory}

In this section we set up the framework for our computation. The Goldstone 
bosons resulting from the breaking of translation invariance are represented by 
a $(d-2)$-component real-valued scalar field $\h (x,t)$ living in a 
2-dimensional rectangular base-space $(x,t)$ of size $r\times \beta$. 
The field $\h$ describes the displacement of the vibrating string from the 
minimal length arrangement in the transverse $(d-2)$ dimensions. The
leading-order free-string approximation is given by the following action
containing two derivatives
\begin{equation}\label{freeaction}
S_2[\h] = \frac{\sigma}{2} \int_0^\beta dt \int_0^r dx \;
\partial_\mu \h \cdot \partial_\mu \h, \quad \mu \in \{x,t\},
\end{equation}
where $\sigma$ is the string tension. In the effective theory the string 
tension is a low-energy parameter whose explicit value can be obtained only
from the underlying Yang-Mills theory. 

In our computation we consider toroidal boundary conditions, i.e.
\begin{equation}
\h(x,t) = \h(x,t + \beta); \quad  
\h(x,t) = \h(x + r,t),
\end{equation}
as well as cylindrical boundary conditions, i.e.
\begin{equation}
\h(x,t) = \h(x,t + \beta); \quad  
\h(0,t) = \h(r,t) = \vec 0,
\end{equation}
for the string. When one interprets the $t$-direction as Euclidean time,
cylindrical boundary conditions describe the propagation of an open string at
finite temperature. On the other hand, when one interprets the $x$-direction as
Euclidean time, the same boundary conditions correspond to the propagation of a 
closed string that is created at ``time'' $x = 0$ and annihilated at $x = r$.
This dual interpretation gives rise to an open-closed string duality of the
string theory.

The first bulk correction to the free-string action contains four derivatives and is 
given by
\begin{equation}\label{NLOaction}
S_4[\h] = \sigma \int_0^\beta dt \int_0^r dx \;
\left[ c_2 (\partial_\mu \h \cdot \partial_\mu \h)^2 +
c_3 (\partial_\mu \h \cdot \partial_\nu \h)^2 \right].
\end{equation}
The open-closed string duality \cite{Lue04} constrains the value of the two 
low-energy parameters $c_2$ and $c_3$ by
\begin{equation}\label{OCduality}
(d-2) \, c_2 + c_3 = \frac{d-4}{8}~.
\end{equation}
Note that, for $d=3$, i.e.\ for a 1-component field $h(x,t)$, the two terms in 
eq.~(\ref{NLOaction}) are identical and hence the first correction is completely 
fixed. As shown in \cite{aha09}, for general $d$ a generalization of L\"uscher 
and Weisz's argument provides a further constraint on the next-to-leading order 
coefficients 
\begin{equation}\label{aha}
c_2 + c_3 = - \frac{1}{8}~.
\end{equation}
Thus, for any value of $d$, the two independent constraints eq.~(\ref{OCduality})
and eq.~(\ref{aha}) completely fix the effective action at this perturbative 
order with $c_2 = \frac{1}{8}$ and $c_3 = - \frac{1}{4}$. Interestingly, the 
expansion of the Nambu-Goto action 
\begin{equation}\label{NGaction}
S_{NG}[\h] = \sigma \int_0^\beta dt \int_0^r dx \;
\sqrt{1 + 
\partial_x\h \cdot \partial_x \h+\partial_t\h \cdot \partial_t \h
+(\partial_x\h\times\partial_t\h)^2  }
\end{equation}
satisfies these two constraints.

Since cylindrical boundary conditions explicitly break translation invariance 
in the $x$-direction, one would expect surface terms (located at the boundaries 
at $x = 0$ and $x = r$) to appear in the effective action. Remarkably, as was 
shown by L\"uscher and Weisz, due to open-closed string duality such terms are 
absent at leading order (i.e.\ there are no boundary terms with two 
derivatives). Boundary terms with an odd number of derivatives are excluded by
parity symmetry. However, boundary terms with four derivatives do indeed exist.
Fortunately, such terms contribute at one order higher than the four-derivative
terms in the bulk that we discussed before. As a result, in our study boundary
terms need not be taken into account.

The squared width of the string is defined as the second moment of the field 
$\h$, i.e.
\begin{equation}
w^2(x,t) = \langle (\h(x,t) - \h_0)^2 \rangle = 
\frac{\int {\cal{D}}\h\; (\h(x,t) - \h_0)^2 \exp(- S[\vec h])}
{\int {\cal{D}}\h\; \exp(- S[\vec h])}.
\end{equation}
Here $S[\h]$ is the effective string action and 
\begin{equation}
\h_0 = \frac{1}{\beta r} \int_0^\beta dt \int_0^r dx \ \h(x,t)
\end{equation} 
is the equilibrium 
position of the string. For cylindrical boundary conditions, we have $\h_0 = 0$.
At next-to-leading order, the string action is given by $S[\h] = S_2[\h] +
S_4[\h]$. Similarly, at next-to-leading order, the field is replaced by
$\h(x,t) \longrightarrow \h(x,t) + \alpha \partial_\mu \partial_\mu \h(x,t)$,
where $\alpha$ is a low-energy parameter. Expanding around the free-string 
action, the squared width of the string is given by
\begin{eqnarray}
w^2(x,t)&=&w_{lo}^2(x,t) - \langle (\h(x,t)^2 - \h_0^2)\, S_4 \rangle_0
+ 2 \alpha\, \langle (\partial_\mu \h(x,t))^2 \rangle_0 \nonumber \\
&+&\alpha^2\, \langle (\partial_\mu \partial_\mu \h (x,t))^2\rangle_0
- \frac{2 \alpha}{\beta r} \int dt \ dx \ \langle \h_0 \cdot 
\partial_\mu\partial_\mu \h(x,t) \rangle_0 \nonumber \\
&-&\frac{\alpha^2}{(\beta r)^2} \int dt \ dx \ dt' \ dx' \ 
\langle \partial_\mu \partial_\mu \h(x,t) \cdot \partial_{\mu'} \partial_{\mu'}
\h(x',t') \, \rangle_0.
\end{eqnarray}
Here $\langle \dots \rangle_0$ represents the vacuum expectation value with
respect to the free-string action and
\begin{equation}
w_{lo}^2(x) = \langle \h(x,t)^2 \rangle_0 - \langle \h_0^2 \rangle_0
\end{equation}
is the result for the squared width at leading order.

Let us define $G(x,t;x',t') = \langle h^a(x,t) h^a(x',t') \rangle_0$ as
the free field propagator of a single component $a$. One then obtains
\begin{equation}\label{NLwidth}
\langle \h(x,t)^2\, S_4 \rangle_0 = 4 \,(d-2)
\left\{\,  \left[ (d-2) c_2 + c_3 \right] T_1 \,+\, \left[ 2 c_2 + (d-1) c_3 
\right] T_2 \, \right\}~.
\end{equation}
Using the two constraints eq.~(\ref{OCduality}) and eq.~(\ref{aha}) this implies
\begin{equation}
\bra \h(x,t)^2\, S_4 \ket_0 = \frac{(d-2)^2}2(T_1-2T_2)-(d-2)T_1.
\label{nlw}
\end{equation}
The two terms $T_1$ and $T_2$ are given by
\begin{equation}\label{T1}
T_1 = \lim _{\epsilon,\, \epsilon'  \rightarrow 0}
\int_0^\beta dt' \int_0^r dx' \;
\partial_{\mu'} G(x,t;x',t') \; \partial_{\mu''} G(x'',t'',x,t) \;  
\partial_{\nu'} \partial_{\nu''} G(x',t';x'',t''),
\end{equation}
as well as
\begin{equation}\label{T2}
T_2 = \lim _{\epsilon,\, \epsilon'  \rightarrow 0}
\int_0^\beta dt' \int_0^r dx' \;
\partial_{\mu'} G(x,t;x',t') \; \partial_{\nu''} G(x'',t'';x,t) \;  
\partial_{\mu'} \partial_{\nu''} G(x',t',x'',t''),
\end{equation}
where $x'' = x' + \epsilon$ and $t'' = t' + \epsilon'$. Since they are 
ultraviolet divergent for $(x',t') = (x'',t'')$, the integrals defined 
above have been regularized using the point-splitting method. Finally, we have
\begin{eqnarray}
&&\langle \h(x,t)\cdot  \partial_{\mu'} \partial_{\mu'} \h(x',t') \rangle_0 = 
(d-2) \, \partial_{\mu'} \partial_{\mu'} G(x,t;x',t'), \nonumber \\
&&\langle \partial_{\mu} \partial_{\mu} \h(x,t) \cdot
\partial_{\nu'} \partial_{\nu'} \h(x',t') \rangle_0 =
(d-2)\, \partial_{\mu} \partial_{\mu}  \partial_{\nu'} \partial_{\nu'} 
G(x,t;x',t').
\end{eqnarray}

\section{Toroidal boundary conditions}

In this section we present the computation of the string width with toroidal 
boundary conditions at next-to-leading order in the low-energy effective 
theory. As we show in Appendix \ref{apptorus}, the single-component free field 
propagator can be written as
\begin{equation}\label{Gtorus}
G(x,t) = \frac{t (t-\beta)}{2 \sigma \beta r} + \frac{1}{2 \pi \sigma} 
\sum_{n=1}^\infty \cos\left(\frac{2 \pi n x}{r}\right) 
\frac{\e^{- 2 \pi n t/r} + q^n\, \e^{2 \pi n t/r}}{n (1 - q^n)} + K,
\end{equation}
where
\begin{equation}
u=\frac{\beta}{r}, \qquad q=\e^{-2 \pi u}, \qquad 
K = \frac{\beta}{12 \sigma r} + \frac{1}{\pi \sigma} \log\eta(iu).
\end{equation}
It should be noted that $t\in [0,\beta]$. In eq.~(\ref{Gtorus}) we have 
used translation 
invariance, i.e.\ $\bra h(x,t)\,h(x',t')\ket = G(x - x',t - t')$. Translation 
invariance also implies that the string width $w(x) = w$ does not 
depend on the 
position $x$. At leading order, the squared width, $w_{lo}^2$, is ultraviolet 
divergent and we regularize it using the point-splitting method
\begin{equation}
w_{lo}^2 = \lim_{\epsilon,\, \epsilon'  \rightarrow 0}
\langle \h(x,t)  \h(x',t') \rangle_0 - \langle \h_0^2 \rangle_0 =
(d-2) \left\{ G(\epsilon,\epsilon') - \int_0^\beta \frac{dt}{\beta} 
\int_0^r \frac{dx}{r} \; G(x,t) \right\}.
\end{equation}
Using eq.~(\ref{tA}) one immediately obtains
\begin{equation}
G(\epsilon,\epsilon') = \frac{1}{2 \pi \sigma} \log\frac{r}{r_0},
\quad r_0 = 2 \pi \sqrt{\epsilon^2 + \epsilon'^2}
\end{equation} 
as well as
\begin{equation}
\int_0^\beta \frac{dt}{\beta} \int_0^r \frac{dx}{r} \; G(x,t) =
- \frac{\beta}{12 \sigma r} + K,
\end{equation}
such that
\begin{equation}
w_{lo}^2 = \frac{d-2}{2 \pi \sigma}\log\frac{r}{r_0} 
- \frac{d-2}{\pi\sigma} \log \eta (iu).
\end{equation}
The quantity $r_0$ is a low-energy parameter of dimension [length]. 

Let us now consider the corrections to this behavior resulting from the 
next-to-leading term, $S_4$, of the effective string action. By explicit 
calculation, it turns out that $\langle \h_0^2\, S_4 \rangle_0 = 0$. Hence, it
remains to evaluate eq.~(\ref{nlw}) with $T_1$ and $T_2$ given by eq.~(\ref{T1}) 
and eq.~(\ref{T2}). We find
\begin{eqnarray}
&&T_1= - \frac{1}{2 \pi \sigma^2\, \beta r} \left[\log\frac{r}{r_0} - 
2 \log\, \eta(iu)\right], \nonumber \\
&&T_2 = \frac{T_1}{2} + \frac{\pi u E_2^2(iu)}{72 \, \sigma^2 r^2}
- \frac{E_2(iu)}{12\, \sigma^2 r^2} + \frac{1}{8 \pi \, \sigma^2 \beta r}.
\end{eqnarray}
Furthermore, using the two identities
\begin{eqnarray}
&&\partial_x \partial_x G(x - x',t - t') = - \partial_t \partial_t 
G(x - x',t - t') + \frac{1}{\sigma \beta r}, \nonumber \\
&&\partial_{x'} \partial_{x'} \partial_x \partial_x G(x - x',t - t') =
\partial_{t'} \partial_{t'} \partial_t \partial_t G(x - x',t - t'),
\end{eqnarray}
we find that the terms proportional to $\alpha$ cancel and those proportional to
$\alpha^2$ vanish. Hence, at next-to leading order the squared width of the 
string is given by
\begin{equation}
w^2 = \left(1 - \frac{1}{\sigma \, \beta r}\right) w_{lo}^2 +
\frac{(d-2)^2}{4 \, \sigma^2 \beta r} \left(\frac{\pi}{18} [u E_2(iu)]^2 -
\frac{u E_2(iu)}{3} + \frac{1}{2 \pi} \right).
\end{equation}
It should be noted that, order by order, this expression is modular invariant. 
In particular, it is invariant under the interchange of $r$ and $\beta$.

\section{Cylindrical boundary conditions}

Together with Appendix \ref{next}, this section contains the calculation 
of the string width at next-to-leading order for cylindrical boundary 
conditions. In Appendix \ref{appcyl}, we show that the single-component 
free field propagator can be written as
\begin{equation}\label{Gcyl}
G(x,t;x',t') = \frac{1}{\pi \sigma} \sum_{n=1}^\infty
\sin\left(\frac{n \pi x}{r}\right) \; \sin\left(\frac{n \pi x'}{r}\right)  
\frac{\e^{- n \pi (t-t')/r} + q^n \; \e^{n \pi (t-t')/r}}{n (1 - q^n)},
\end{equation}
with
\begin{equation}
u = \frac{\beta}{2r}, \quad q = \e^{-2 \pi u}.
\end{equation}
In eq.~(\ref{Gcyl}) we have used that $(t-t')\in [0,\beta]$. We now calculate 
the string width $w(r/2)$ at the midpoint $x = r/2$. It should be noted that, 
due to translation invariance in the $t$-direction, the string width does not 
depend on $t$. At leading order, the squared width $w_{lo}^2(r/2)$ is 
ultraviolet divergent and is again regularized using the point-splitting 
method. It turns out that
\begin{equation}
w_{lo}^2(r/2) = \frac{d-2}{2 \pi \sigma} \log\frac{r}{r_0} +
\frac{d-2}{\pi \sigma} \log\frac{\eta(2i u)}{\eta^2(iu)},
\label{w0C}
\end{equation}
where the low-energy parameter $r_0$ is now given by
\begin{equation}
r_0 = \frac{\pi}{2} \sqrt{\epsilon^2 + \epsilon'^2}.
\label{rcyl}
\end{equation} 
For $\beta \gg r$, the second term on the right-hand side of eq.~(\ref{w0C}) gives only
exponentially small corrections to the leading logarithmic increase of the string
width. The regime $r \gg \beta$~\cite{All08} can be obtained using the inversion
transformation rule given by eq.~(\ref{invrl}). Then we have  
\begin{equation}
w_{lo}^2(r/2) = \frac{d-2}{2 \pi \sigma} \log\frac{\beta}{4 r_0} +
\frac{d-2}{4\beta \sigma}\, r +{\cal{O}}(\mbox{e}^{-2\pi r/\beta}).
\end{equation}
Interestingly, this equation shows that at finite but low temperature, the squared string
width increases linearly with the distance. 
Similar to the case of toroidal boundary conditions, we have to evaluate
eq.~(\ref{nlw}). Since eq.~(\ref{Gcyl}) is well-defined only for 
$(t-t')\in [0,\beta]$, it is convenient to split the integral over $t$, i.e.\
$\int_0^\beta dt = \int_0^{\epsilon'} dt + \int_{\epsilon'}^\beta dt$. We then 
obtain
\begin{eqnarray}
T_1&=&\frac{\pi}{\sigma^2 r^4}
\sum_{n=1}^\infty \sum_{m=1}^\infty (-1)^{m+n} \sum_{k=1}^\infty \frac{k}{1 - q^k}
\left(\e^{- \pi k \epsilon'/r} + q^k \e^{\pi k \epsilon'/r}\right) \nonumber \\
&\times&\int_0^r dx \, \cos\frac{\pi (2 k x + k \epsilon)}{r} 
\Big[\cos\frac{\pi(2 (m-n) x + (2m-1) \epsilon)}{r} \nonumber \\
&\times&\frac{1}{(1-q^{2n-1})(1-q^{2m-1})} 
\int_{\epsilon'}^\beta dt \left(\e^{-2 \pi (n+m-1) t/r} + 
\e^{-2 \pi (n+m-1) (\beta-t)/r}\right) \nonumber \\
&+&\cos\frac{\pi(2 (m+n-1) x + (2m-1) \epsilon)}{r} \nonumber \\
&\times&\frac{1}{(1-q^{2n-1})(1-q^{2m-1})}
\int_{\epsilon'}^\beta dt \left(\e^{- 2 \pi (n-m) t/r} q^{2m-1} +
\e^{- 2 \pi (m-n) t/r} q^{2n-1}\right)\Big],
\label{T1C}
\end{eqnarray}
as well as
\begin{eqnarray}
T_2&=&\frac{T_1}{2}
- \sum_{n=1}^\infty \sum_{m=1}^\infty (-1)^{m+n} \ \frac{E_2(iu)}{24 r^4 \sigma^2}
\int_0^r dx \, \cos\frac{2\pi (m-n) x}{r} \nonumber \\
&\times&\frac{1}{(1-q^{2n-1})(1-q^{2m-1})}
\int_{\epsilon'}^\beta dt \left(\e^{- 2 \pi (n-m) t/r} q^{2m-1} +
\e^{- 2 \pi (m-n) t/r} q^{2n-1}\right) \nonumber \\
&-&\frac{E_2(iu)}{96 \sigma^2 r^2}~.
\label{T2C}
\end{eqnarray} 
The last term is the contribution of the integration $\int_0^{\epsilon'} dt$ 
after the limit $\epsilon' \to 0$ has been taken. The explicit evaluation of 
$T_1$ and $T_2$ is rather involved, and is thus relegated to Appendix 
\ref{next}. 

The two identities
\begin{eqnarray}
&&\partial_x \partial_x G(x,t;x',t') = - \partial_t \partial_t G(x,t;x',t'),
\nonumber \\
&&\partial_{x'} \partial_{x'} \partial_x \partial_x G(x,t;x',t') =
\partial_{t'} \partial_{t'} \partial_t \partial_t G(x,t;x',t'),
\end{eqnarray}
imply that $\langle (\partial_\mu \h(x,t))^2 \rangle_0 = 0$ and 
$\langle (\partial_\mu \partial_\mu \h(x,t))^2\rangle_0 = 0$. Hence the 
contribution to the squared width coming from the terms proportional to 
$\alpha$ and $\alpha^2$ vanishes at next-to-leading order. Finally, inserting 
eq.~(\ref{T1CF}) and eq.~(\ref{T21}) in eq.~(\ref{nlw}) we
obtain
\begin{eqnarray}
w^2(r/2)&=&w^2_{lo}(r/2) + \frac{\pi}{12 \sigma r^2}
\left[E_2(iu) - 4 E_2(2iu)\right]
\left(w_{lo}^2(r/2) - \frac{d-2}{4 \pi \sigma}\right) \nonumber \\
&+&\frac{(d-2) \pi}{12 \sigma^2 r^2} \Big\{ 
u \left(q \frac{d}{dq} - \frac{d-2}{12} E_2(iu)\right)
\left[E_2(2iu) - E_2(iu)\right] - \frac{d-2}{8 \pi} E_2(iu)\Big\}.\qquad
\label{w2C}
\end{eqnarray}

\section{Conclusions}

By now, a lot of numerical evidence supporting the validity of the low-energy
effective string description of the long-distance static quark-anti-quark 
potential has been accumulated in lattice Yang-Mills theory. In several cases,
the numerical data are so accurate that higher-order corrections to the leading
free string approximation must be taken into account. Remarkably, open-closed 
string duality completely determines the terms in the effective action at 
next-to-leading order, without any additional low-energy parameters. In this 
paper we have presented the details of an analytic computation of the width of 
the color flux tube in the low-energy effective theory at next-to-leading 
order. Our result has been crucial for accurately describing the width of the 
color flux tube obtained in numerical simulations of Yang-Mills theory 
\cite{gpw}. The results are expressed in closed form in terms of the 
Dedekind~$\eta$ function and of the Eisenstein series $E_2$ and $E_4$. The 
modular inversion transformation property of those functions yields the 
next-to-leading order correction to the width both at zero and at finite (but
low) temperatures. The calculation has been performed using both toroidal and 
cylindrical boundary conditions. 

The effective theory that we used describes string fluctuations in the 
continuum. In order to apply the results of our calculation to lattice field
theories, one must be sufficiently close to the continuum limit. Before one 
reaches the continuum limit, the confining string in a lattice Yang-Mills 
theory is also affected by lattice artifacts. First of all, at very strong 
coupling the world-sheet swept out by the lattice string is rigid, i.e.\ it 
follows the discrete lattice steps and does not even have massless excitations. 
Only at weaker coupling, after crossing the roughening transition, the string 
world-sheet supports massless excitations and thus becomes rough. Consequently, 
the effective theory is applicable only in the rough phase. Since the lattice 
theory is invariant only under discrete rotations and not under the full 
Poincar\'e group, before one reaches the continuum limit additional terms 
proportional to $\sum_{\mu = 1,2} (\partial_\mu \partial_\mu h)^2$ and 
$\sum_{\mu = 1,2} (\partial_\mu h)^4$ enter the effective action in the bulk. Since 
these terms contain four derivatives, they are of sub-leading order. Hence,
they have no effect on the L\"uscher term or on the leading logarithmic 
behavior of the string width. As a result, the L\"uscher term is completely 
universal. Provided its world-sheet is rough, even a lattice string supports 
exactly massless modes which contribute $- \pi/24 r$ to the static quark
potential. In order to incorporate lattice artifacts in the effective theory in
a systematic manner, one must investigate whether additional boundary terms 
arise in the effective action. One must also reinvestigate the consequences of
open-closed string duality, which were derived in [20] assuming full Poincar\'e
invariance. These are interesting problems for future studies, which may
eventually be important for the correct description of numerical simulation
data when the lattice spacing is not sufficiently small.

\acknowledgments M.\ P.\ and U.-J.\ W.\ gratefully acknowledge helpful discussions with
P.\ Hasenfratz, F.\ Niedermayer, R.\ Sommer, and P.\ Weisz. We like to thank the anonymous
referee for very insightful remarks and useful suggestions.  This work is supported in
part by funds provided by the Schweizerischer Nationalfonds (SNF). The ``Albert Einstein
Center for Fundamental Physics'' at Bern University is supported by the ``Innovations- und
Kooperationsprojekt C-13'' of the Schweizerische Uni\-ver\-si\-t\"ats\-kon\-fe\-renz
(SUK/CRUS).

\appendix

\section{Infinite sums and products}
\label{appspecfun}

In this appendix we list the infinite sums and products which appear in our 
calculation. Some of them can be expressed in terms of the Dedekind $\eta$ 
function and the Eisenstein series. They are respectively defined by 
\begin{equation}
\eta(\tau)=q^{\frac1{24}}\prod_{n=1}^\infty(1-q^n)~,~ q = \e^{i2\pi\tau}
\label{eta}
\end{equation}
and
\begin{equation}
E_{2k}(\tau) = 1 + (-1)^k \frac{4k}{B_k} \sum_{n=1}^\infty
\frac{n^{2k-1}q^n}{1-q^n}~,
\label{e2k}
\end{equation}
where $B_k$ are the Bernoulli numbers, defined through the expansion
\begin{equation}
\frac{z}{\e^z - 1} = 1 - \frac{z}{2} - 
\sum_{k=1}^\infty(-1)^k \frac{B_k}{(2k)!} z^{2k}~.
\label{bernoulli}
\end{equation}
They are also related to the Riemann $\zeta$ function of positive even 
integers and negative odd integers
\begin{equation}
\zeta(2k) = \frac{B_k(2 \pi)^{2k}}{(2k)!}\,,
\qquad
\zeta(1-2k) = (-1)^k \frac{B_k}{2k}~.
\end{equation}
In the one-loop calculation one encounters the sum
\begin{equation}
\sum_{n=1}^\infty \frac{n^{-1} q^n}{1 - q^n} \equiv
\sum_{k=1}^\infty \sum_{n=1}^\infty
\frac{q^{k\,n}}{n} = - \sum_{k=1}^\infty \log\left(1 - q^k\right) = 
- \log\varphi(\tau)~,
\label{id}
\end{equation}
where $\varphi(\tau)$ is the Euler function, related to $\eta$ by
\begin{equation}
\eta(\tau)=q^{1/24}\varphi(\tau)~.
\end{equation}
In two-loop calculations one finds sums of the type  
\begin{equation}
\sum_{n=1}^\infty \frac{nq^n}{1-q^n} \equiv
\sum_{k=1}^\infty \frac{q^k}{(1 - q^k)^2} = 
\frac{1}{24} \left[1 - E_2(\tau)\right]~,
\label{id2}
\end{equation}
where the first identity is simply obtained by writing 
$q^n/(1 - q^n) = \sum_{k=1}^\infty q^{n\,k}$ (like in eq.~(\ref{id})) and then 
inverting the order of the two sums. The second identity follows from the 
definition of $E_2(\tau)$ or from eq.~(\ref{id}) using the relation 
\begin{equation}
\frac{\eta'(\tau)}{\eta(\tau)} = \frac{i\pi}{12} \, E_2(\tau)~.
\end{equation}
The identity (\ref{id2}) also implies the following three equations:
\begin{equation}
\sum_{m=1}^\infty \frac{q^{2m-1}} {(1 - q^{2m-1})^2} = 
\frac{1}{24} \left[E_2(2\tau) - E_2(\tau)\right]~,
\label{id3}
\end{equation}
as well as 
\begin{equation}
\sum_{j=1}^\infty \frac{q^j}{(1 + q^j)^2} = - \frac{1}{8} + \frac{E_2(2\tau)}{6}
- \frac{E_2(\tau)}{24}~,
\label{id6}
\end{equation}
and
\begin{equation}
\sum_{k=1}^\infty \frac{(2m - 1)\,q^{2m-1}(1 + q^{2m-1})}{(1-q^{2m-1})^3} =
\frac{q}{24} \frac{d}{dq} \left[E_2(2\tau)-E_2(\tau)\right]~.
\label{id4}
\end{equation}
The derivative of $E_2(\tau)$ can be expressed in terms of other Eisenstein 
series using
\begin{equation}
q \frac{d}{dq} E_2(\tau) = \frac{1}{12} \left[E_2(\tau)^2 - E_4(\tau)\right]~.
\label{id5}
\end{equation}
The Dedekind $\eta$ function and the Eisenstein series $E_{2k}$ obey the 
following transformation rules under the inversion $\tau \to -1/ \tau$ 
\begin{equation}\label{invrl}
\eta(\tau) = \frac{1}{\sqrt{-i\tau}} \eta(\frac{-1}{\tau})\,,\qquad
E_2(\tau) = \frac{1}{\tau^2} \, E_2(\frac{-1}{\tau}) - \frac{6}{i \pi \tau}~.
\end{equation}
as well as
\begin{equation}
E_{2k}(\tau) = \frac{1}{\tau^{2k}} \, E_{2k}(\frac{-1}{\tau})~,~~ k > 1 ~.
\end{equation}
These relations ensure fast convergence of the expressions in both the regimes
$\beta \gg r \gg 1/\sqrt{\sigma}$ (the zero-temperature limit) and 
$r \gg \beta \gg 1/\sqrt{\sigma}$ (the finite (but low) temperature case).

\section{Propagator on the torus}
\label{apptorus}

The free field two-point function on a torus of size $r\times\beta$ satisfies 
\begin{equation}
- \Delta G(x,t;x't') = \frac{1}{\sigma} \delta(x-x')\,\delta(t-t') - 
\frac{1}{\sigma \, \beta \, r}~.
\end{equation} 
Here the term $1/\sigma\, \beta\, r$ accounts for the zero-mode subtraction, 
which makes the Laplace operator $\Delta$ invertible in the orthogonal 
subspace. The solution of this equation can be expressed in closed form in 
various ways. We begin with the explicit expression given in Appendix A of 
\cite{cas96} 
\begin{equation}
G(z)= - \frac{1}{2 \pi \sigma}
\Re e \left[\log \frac{2\pi z}{r} - \sum_{k=1}^\infty 
\frac{G_k(2\pi)^{2k}}{2k}\left(\frac{z}{r}\right) ^{2k} \right] +
\frac{(t-t')^2}{2 \, \sigma \, \beta \, r}
\label{tA}
\end{equation}
with
\begin{equation}
z = x - x' + i (t - t')~,~G_k = 2 \frac{\zeta(2k)}{r^{2k}}\,E_{2k}(iu)~,~~
u = \frac{\beta}{r}~,
\end{equation}
where the Eisenstein series $E_{2k}(\tau)$ are defined in our appendix
\ref{appspecfun}. Inserting this definition in eq.~(\ref{tA}) we get 
\begin{eqnarray}
2 \pi \sigma G(z)&=&\frac{\pi(\Im m z)^2}{\beta \, r}
- \Re e\left[\log \frac{2\pi z}{r} - \sum_{k=1}^\infty \frac{B_k (2\pi)^{2k}}{(2k)!}
\frac{1}{2k} \left(\frac{z}{r}\right)^{2k}\right] \nonumber \\
&+&2\,\Re e \left[\sum_{k=1}^\infty \frac{(-1)^k}{(2k)!}
\left(\frac{2 \pi z}{r}\right)^{2k} \sum_{n=1}^\infty
\frac{n^{2k-1} q^n}{1 - q^n}\right]~.
\label{tB}
\end{eqnarray}
The terms enclosed in the first square brackets can be re-summed using 
eq.~(\ref{bernoulli})
\begin{equation}
\sum_{k=1}^\infty (-1)^k \frac{B_k\,w^{2k}}{(2k)!}\frac{1}{2k} = \sum_{n=1}^\infty
\frac{\e^{-n\,w}}{n} - \frac{w}{2} + \log w = - \log(1 - \e^{-w}) - 
\frac{w}{2} + \log w,
\label{ber}
\end{equation}
with $w = - 2\pi i z/r$. Interchanging the order of the two sums in the last 
bracket of eq.~(\ref{tB}), we recognize the Taylor expansion of the cosine and 
arrive at the simpler expression
\begin{eqnarray}
2 \pi \sigma G(z)&=&\frac{\pi (t - t')(t - t' - \beta)}{\beta \, r} \nonumber \\
&+&\Re e \left[- \log(1 - \e^{2 \pi i z/r})
+ 2 \sum_{n=1}^\infty \frac{n^{-1}q^n}{1 - q^n}
\cos\frac{2 \pi n z}{r} + 2 \log\varphi(iu)\right]~,
\label{torus}
\end{eqnarray}
where the Euler function $\varphi$ has already been defined in Appendix 
\ref{appspecfun}. Using the identity (\ref{id}) and expanding the first 
logarithm in powers of $\e^{2 \pi i z/r}$, we obtain the Gaussian correlator on 
the torus in the compact form of eq.~(\ref{Gtorus}), where we have put 
$G(x,t) = G(z)$. It should be noted that this expression converges in the range 
$q<1$ and $0 \le t - t' \le \beta$.

\section{Propagator on the cylinder}
\label{appcyl}

The Gaussian correlator $G(x,t;x',t') \equiv \bra h(x,t)\,h(x',t') \ket_0$ on a 
cylinder of size $r \times \beta$ with fixed boundary conditions at $x = 0$ and $x = r$ 
and periodic boundary conditions in $t$ with period $\beta$ can be conveniently 
written in terms of correlators $G(z - z')$ and $G(z + z'^*)$ on a torus of 
size $2 r \times \beta$ (with $z' = x' + i t'$ and $z'^* = x' - i t'$)
\begin{eqnarray}
G(x,t;x',t')&=&G(z - z') - G(z + z'^*) \nonumber \\
&=&\frac{1}{\pi \sigma} \sum_{n=1}^\infty \sin\frac{\pi n x}{r} \;
\sin\frac{\pi n x'}{r}
\frac{\e^{-\pi n (t - t')/r} + \e^{- \pi n (\beta - t + t')/r}}{n (1 - q^n)}~.
\label{cylinder}
\end{eqnarray}
This is the expression used in section 4.

\section{Evaluation of $T_1$ and $T_2$ on the cylinder}
\label{next}

In this appendix we evaluate explicitly the terms $T_1$ and $T_2$ given in 
eq.~(\ref{T1C}) and eq.~(\ref{T2C}). Performing the corresponding integrations 
one obtains
\begin{equation}
T_2 - \frac{1}{2} T_1 =
\frac{\pi}{\sigma^2 r^2} \ \frac{\beta}{r} \, \sum_{m=1}^\infty
\left[- \frac{1}{12} E_2(iu) \frac{q^{2m-1}}{(1 - q^{2m-1})^2}\right] - 
\frac{1}{\sigma^2 r^2} \frac{E_2(iu)}{96}~.
\end{equation}
Using eq.~(\ref{id3}), the sum may be written in closed form, which then yields
\begin{equation}
T_2 - \frac{1}{2} T_1 = - \frac{\pi}{\sigma^2 r^2} 
\frac{u\,E_2(iu) \left[E_2(2iu) - E_2(iu)\right]}{144} - 
\frac1{\sigma^2 r^2} \frac{E_2(iu)}{96}~.
\label{T21}
\end{equation}
Performing the integrations in $T_1$ and putting 
$s = \e^{- \pi \epsilon'/r}$, we obtain
\begin{eqnarray}
\label{tsum}
T_1&=&\frac{1}{\sigma^2 r^2} \sum_{k=1}^\infty \sum_{n=1}^\infty (-1)^k
\Big[\frac{k}{k+2n-1} \cos\frac{\pi(k + 2n - 1)\epsilon}{R} \ 
\frac{s^k + q^k}{1 - q^k} \nonumber \\
&\times&\Big(\frac{s^{2k+2n-1} + s^{2n-1}}{2} +
\frac{q^{2k+2n-1}}{1 - q^{2k+2n-1}} + \frac{q^{2n-1}}{1 - q^{2n-1}}\Big) 
\nonumber \\
&-&\frac{k+2n-1}{k}\cos\frac{\pi k \epsilon}{r} \ 
\frac{s^{k+2n-1} + q^{k+2n-1}}{1 - q^{k+2n-1}}
\Big(\frac{q^{2k+2n-1}}{1 - q^{2k+2n-1}} - 
\frac{q^{2n-1}}{1 - q^{2n-1}}\Big)\Big] \nonumber \\
&+&\frac{\pi}{\sigma^2 r^2} \ \frac{\beta}{r} \,\sum_{m=1}^\infty
\Big[\frac{(2m-1) \, q^{2m-1}(1 + q^{2m-1})}{(1 - q^{2m-1})^3}\Big].
\end{eqnarray}
The last sum may be written in terms of Eisenstein series using eq.~(\ref{id4}) 
or eq.~(\ref{id5}).

Introducing the two functions
\begin{equation}
A(a,b)= \sum_{k=1}^\infty \sum_{n=1}^\infty (-1)^k \frac{k}{k+2n-1} a^k 
b^{k+2n-1}~,
\end{equation}
as well as
\begin{equation}
B(a,b)= \sum_{k=1}^\infty \sum_{n=1}^\infty (-1)^k \frac{k + 2n - 1}{k} a^k 
b^{k+2n-1}~,
\end{equation}
and putting $t = \e^{i \pi (\epsilon + i\epsilon')/r}$, we can rewrite 
eq.~(\ref{tsum}) in the form
\begin{eqnarray}
T_1&=&\frac{1}{\sigma^2 r^2} \Re\,e \Big\{\frac{A(1,t)+A(s^2,t)}{2} \nonumber \\
&+&\sum_{j=1}^\infty \big[A(q^j,q^j) + A(q^{-j},q^j) + 2 \, A(q^j,t)
- B(q^j,s \, q^j) + B(q^{-j},s \, q^j) + 2\,A(1,q^j) \nonumber \\
&+&2 \,\sum_{k=1}^\infty \left[A(q^{j+k},q^j) - B(q^j,q^{j+k}) + 
B(q^{-j},q^{j+k})\right] + 2 \sum_{k=1,k\not=j}^\infty A(q^{k-j},q^j) \big]\Big\}
\nonumber \\
&+&\frac{\pi\,u}{12 \sigma^2 r^2} \,
q \frac{d}{dq} \left[E_2(2iu) - E_2(iu)\right].
\label{ng1}
\end{eqnarray}
The two functions $A$ and $B$ can be written in terms of elementary functions
\begin{eqnarray}
&&A(a,b) = -\frac{a^2 \, b}{(1 - a^2)(1 + a b)}
- \frac{a(1 + a^2)}{(1 - a^2)^2} \log(1 + a b)
+ \frac{a \log(1 - b)}{2(1 + a)^2} + \frac{a \log(1 + b)}{2(1 - a)^2}~,
\nonumber \\
&&B(a,b) = -\frac{a \, b^2}{(1 - b^2)(1 + a b)} -
\frac{b(1 + b^2)}{(1 - b^2)^2} \log(1 + a b)~.
\end{eqnarray}
The divergences of $A(a,b)$ for $a \to 1$ or $b \to 1$ are only apparent. In 
particular, the useful ultraviolet limits to apply in eq.~(\ref{ng1}) are 
\begin{eqnarray}
\label{a1q}
&&A(1,q) = \frac{q}{4(1 + q)^2} + \frac{1}{8} \log\frac{1 - q}{1 + q}~, 
\nonumber \\
&&A(q,t) = q \frac{- q(1 - q) + (1 + q^2) \log\frac{2}{1 + q}}{(1 - q^2)^2} -
\frac{q}{2(1 + q)^2} \log\frac{2r}{\pi \vert\epsilon + i \epsilon'\vert}
+ {\cal O}(\epsilon + i\epsilon')~, \nonumber \\
&&\Re\,e \left[\frac{A(1,t) + A(s^2,t)}{2}\right] = \frac{1}{16} -
\frac{1}{8} \log\frac{2r}{\pi \vert\epsilon + i \epsilon'\vert} +
{\cal O}(\epsilon + i\epsilon')~.
\end{eqnarray}
We now rewrite this quantity in terms of Eisenstein series. We begin by 
explicitly writing the logarithmic terms of the double sum in eq.~(\ref{ng1}), 
namely those arising from
\begin{equation}
\sum_{j=1}^\infty \Big\{2 \, \sum_{k=1}^\infty \left[A(q^{j+k},q^j)
- B(q^j,q^{j+k}) + B(q^{-j},q^{j+k}) \right] + 2 \sum_{k=1,k\not=j}^\infty 
A(q^{k-j},q^j)\Big\}~, 
\end{equation}
which are given by
\begin{eqnarray}
\label{jk}
&&\sum_{j=1}^\infty \Big\{\sum_{k=1}^\infty \Big[
\frac{q^{j+k}}{(1 + q^{j+k})^2} \log(1 - q^k) +
\frac{q^{j+k}}{(1 - q^{j+k})^2} \log(1 + q^k) \nonumber \\
&+&\frac{q^{j-k}}{(1 + q^{j-k})^2} \log(1 - q^k) 
- 2 \frac{q^{j+k} (1 + q^{2(j+k)})}{(1 - q^{2(j+k)})^2} \log(1 + q^k)\Big] 
\nonumber \\
&+&\sum_{k=1,k\not=j}^\infty \Big[\frac{q^j \log(1 + q^k)}{q^k(1 - q^{j-k})^2} -
2 \frac{q^k(1 + q^{2(k-j)})}{q^j(1 - q^{2(k-j)})^2} \log(1 + q^k)\Big]\Big\}
\nonumber \\
&&= \sum_{k=1}^\infty \log\frac{1 - q^k}{1 + q^k} \left[\sum_{j=1}^\infty
\frac{q^{j+k}}{(1 + q^{j+k})^2} + \sum_{j=1,j\not=k}^\infty
\frac{q^{j-k}}{(1 + q^{j-k})^2}\right] \nonumber \\
&&= 2 \sum_{m=1}^\infty \frac{q^{m}}{(1 + q^{m})^2} \sum_{k=1}^\infty
\log\frac{1 - q^k}{1 + q^k} - \sum_{j=1}^\infty \frac{q^j}{(1 + q^j)^2}
\log\frac{1 - q^j}{1+q^j}~.
\end{eqnarray}
The last sum cancels exactly against the logarithmic terms resulting from the 
single sums (i.e.\ the second line) in eq.~(\ref{ng1}). Apart from the first 
term in eq.~(\ref{jk}), the only remaining terms are those associated with 
eq.~(\ref{a1q}). Putting all these terms together we obtain
\begin{eqnarray}
\label{T1C1}
T_1&=&\frac{1}{\sigma^2 r^2} \left(\sum_{j=1}^\infty \frac{q^j}{(1 + q^j)^2} +
\frac{1}{8}\right) \left[\frac{1}{2} - \log\left(\frac{2r}
{\pi\vert\epsilon + i\epsilon'\vert} \prod_{k=1}^\infty
\frac{(1 + q^k)^2}{(1 - q^k)^2}\right)\right] \nonumber \\
&+&\frac{\pi \, u}{12 \sigma^2 r^2} \,
q \frac{d}{dq} \left[E_2(2iu) - E_2(iu)\right]~.
\end{eqnarray}
By comparison with eq.~(\ref{w0C}) in the second factor of the first term we 
recognize the contribution to the squared width of the flux tube at leading 
order. Applying the identity (\ref{id6}) and the definition (\ref{eta}) of the 
Dedekind $\eta$ function, we finally obtain 
\begin{equation}
\label{T1CF}
T_1 = \frac{1}{24 \sigma^2 r^2} \left\{\left[E_2(iu) - 4 E_2(2iu)\right] 
\left(\log\frac{r \,\eta^2(2iu)}{r_0 \,\eta^4(iu)} - \frac{1}{2}\right) + 
2 \pi u \, q \frac{d}{dq} \left[E_2(2iu) - E_2(iu)\right]\right\},
\end{equation}
where $r_0$ has been defined in eq.~(\ref{rcyl}). Inserting eq.~(\ref{T1CF}) in 
eq.~(\ref{T21}) one immediately obtains an explicit expression for $T_2$ in 
terms of Eisenstein series.


\begin{thebibliography}{20}

\bibitem{Lue80}
M.\ L\"uscher, K.\ Symanzik, and P.\ Weisz, Nucl.\ Phys.\ B173 (1980) 365.

\bibitem{Lue81}
M.\ L\"uscher, Nucl.\ Phys.\ B180 (1981) 317.

\bibitem{Lue81a}
M.\ L\"uscher, G.\ M\"unster, and P.\ Weisz, Nucl.\ Phys.\ B180 (1981) 1.

\bibitem{has}
M.\ Hasenbusch and K.\ Pinn, Physica A 192 (1993) 342.

\bibitem{cas96}
M.\ Caselle, F.\ Gliozzi, U.\ Magnea, and S.\ Vinti, Nucl.\ Phys.\ B460 (1996) 397.

\bibitem{Cas96a}
M.~Caselle, R.~Fiore, F.~Gliozzi, M.~Hasenbusch, and P.~Provero,
Nucl.\ Phys.\  B486 (1997) 245.

\bibitem{Luc01}
B.~Lucini and M.~Teper,
Phys.\ Rev.\  D64 (2001) 105019.

\bibitem{Jug02}
K.~J.~Juge, J.~Kuti, and C.~Morningstar,
Phys.\ Rev.\ Lett.\ 90 (2003) 161601.

\bibitem{Lue02}
M.\ L\"uscher and P.\ Weisz, JHEP 0207 (2002) 049.

\bibitem{Cas04}
M.~Caselle, M.~Pepe, and A.~Rago,
JHEP 0410 (2004) 005.

\bibitem{Cas06}
M.~Caselle, M.~Hasenbusch, and M.~Panero,
JHEP  0603 (2006) 084. 

\bibitem{Har06}
N.~D.~Hari Dass and P.~Majumdar,
JHEP 0610 (2006) 020.

\bibitem{giu07} P. Giudice, F. Gliozzi, and S. Lottini, 
JHEP 0701 (2007) 084.

\bibitem{Bri08}
B.~Bringoltz and M.~Teper,
Phys.\ Lett.\ B 663 (2008) 429.

\bibitem{Ath08}
A.~Athenodorou, B.~Bringoltz, and M.~Teper,
JHEP 0905 (2009) 019.

\bibitem{All08}
A.~Allais and M.~Caselle,
JHEP 0901 (2009) 073.

\bibitem{Bra09}
B.~B.~Brandt and P.~Majumdar,
Phys.\ Lett.\ B682  (2009) 253.

\bibitem{gpw} F.\ Gliozzi, M.\ Pepe, and U.-J.\ Wiese,
Phys.\ Rev.\ Lett.\ 104 (2010) 232001.

\bibitem{Bak10}
A.~S.~Bakry, D.~B.~Leinweber, P.~J.~Moran, A.~Sternbeck, and A.~G.~Williams,
 arXiv:1004.0782 [hep-lat].

\bibitem{Lue04}
M.\ L\"uscher and P.\ Weisz, JHEP 0407 (2004) 014.

\bibitem{aha09}
O.~Aharony and E.~Karzbrun, JHEP  0906 (2009) 012

\end{thebibliography}
\end{document}